# Paramagnetism of layered organic conductors (BEDO-TTF)$_2$ReO$_4$·H$_2$O and (BEDT-TTF)$_x$Dy(NO$_3$)$_z$: analysis of phase transition at 200 K

Yu.N. Shvachko, D.V. Starichenko, A.V. Korolyov,
*Institute of Metal Physics, Ural Division of Russian Academy of Sciences, Ekaterinburg, Russian Federation*

N.D. Kushch, E.B. Yagubskii
*Institute of Problems of Chemical Physics, Russian Academy of Sciences, Chernogolovka, Russian Federation*

Layered organic conductors **(BEDO-TTF)$_2$ReO$_4$·H$_2$O** (metal) and **(BEDT-TTF)$_x$Dy(NO$_3$)$_z$** (paramagnetic insulator) are studied by CW-EPR and SQUID methods. A correlation in resonance and transport properties of both compounds is attributed to a similar mechanism of fine anion ordering (AO) at $T_{AO}$~200 K. Due to the hydrogen bonds motif the small configuration changes in anion sub-lattice bring a profound effect on electronic properties. It is shown that the spin system of **(BEDO-TTF)$_2$ReO$_4$·H$_2$O** consists of a fraction of delocalized π-type holes ($I_{epr}$~$n(\varepsilon_F)$), $I_{epr}$(300K)=1,62·10$^{-4}$ emu/mol and antiferromagnetically correlated local moments S=1/2, $\chi_p$(300K)=1,86·10$^{-3}$ emu/mol. The phase transition Me-Me' at $T_{AO}$=203 K is detected by paramagnetic relaxation and resisitivity but it is not observed in spin susceptibility. Fraction of delocalized holes gradually decreases by factor 2 at cooling down to 100 K, whereas contribution from local moments reaches maximum at ~50 K and falls to zero at 14 K. However, at 2 K total spin susceptibility is fully recovered approaching 6·10$^{-2}$ emu/mol. In turn, **(BEDT-TTF)$_x$Dy(NO$_3$)$_z$** does not have solvent "bridges" (**H$_2$O**) between neighboring anions and the disorder in anion layer converts this compound into paramagnetic insulator. Moreover, the anion metal-complexes **Dy(NO$_3$)$_z$$^{n-}$** contain magnetic ions **Dy$^{3+}$**. Nonetheless, the phase transition of similar origin is also detected at T=197 K by ESR and resistivity measurements. Two spin sub-systems co-exist in this compound: hopping local moments S=1/2 (**BEDT-TTF$^+$**), $I_{epr}$(300K)=6,3·10$^{-4}$ emu/mol and strongly localized 4f$^9$-electrons, J=15/2 (**Dy$^{3+}$**), $\chi_p$(300K)=4·10$^{-2}$ emu/mol. Although no inter-layer (π-f) or intra-layer (f-f) magnetic interactions have been observed within 2÷350 K the **BEDT-TTF**–based conductors with polyhedral metal-complex anions are considered as prospective magneto active materials.

## I. INTRODUCTION

Charge transfer complexes based on BEDT-TTF (bis(ethylenedithio) tetrathiafulvalene) donor molecule or its derivatives demonstrate great variety of crystal structures with two-dimensional conducting sheets alternating by insulating anion layers. The conducting plane consists of stacks of BEDT-TTF cation-radicals with their long axes nearly perpendicular to the plane. Due to spatial proximity, the molecular orbitals of sulfur hetero atoms substantially overlap forming π-type conduction band. Charge transfer induces partial filling of the band allowing hopping holes within the layer and metallic behavior at half filling.

Electric properties of BEDT-TTF-based materials with inorganic tetrahedral, octahedral and polyhedral anions (InBr$_4$, BrO$_4$, GaI$_4$, InI$_4$, CuBr$_4$, AsF$_6$, SbF$_6$, PF$_6$ *etc.*) [1] are characterized by metal-insulator transitions at high temperatures 90-300 K. In most compounds the order-disorder transitions in anion layers are responsible for drastic changes of transport properties. Thus, these materials can be considered as 'organic–inorganic molecular composites' or 'chemically constructed multilayers' [2].

In metallic regime the electronic properties can be considered as two dimensional [3, 4]. In the Mott insulator regime the $\pi$-system can be regarded as having a localized moment $S=1/2$ on each BEDT-TTF di-mer, or n-mer depending on stoichiometry. The characteristics of MI phase transition are very sensitive to the peculiarities of crystal structure and hydrogen bonding with anions. Therefore, the magnetic interactions in anion layer might be important for macroscopic transport properties of the compound at relatively high temperatures.

The idea of introduction of magnetic ions into the anion layer was earlier realized in organic metal (BEDT-TTF)$_3$CuCl$_4\cdot$H$_2$O, where the electrons close to the Fermi surface are largely confined to the frontier orbitals of the BEDT-TTF, while the magnetic moments are localized on the anions, providing a degree of spatial separation between the two [5]. Next step was initiated by discovery of molecular paramagnetic superconductor (BEDT-TTF)$_2$H$_2$OFe(C$_2$O$_4$)$_3$C$_6$H$_5$CN with $T_c$=7 K [6]. Although no long range magnetic ordering was achieved the unusual magnetotransport properties have been recently observed in a series of similar compounds. Note, that the solvent H$_2$O molecules are present in most paramagnetic metals indicating their possible stabilization function in preventing from MI transition. Recent discovery of the first magnetic metal (BEDT-TTF)$_3$[MnCr(C$_2$O$_4$)$_3$] [7] with weak ferromagnetism at $T_c$=5,5 K stimulate interest of researches in synthesis of new hybrid materials with higher $T_c$. Meantime, in current work we attract attention to the influence of magnetic state in anion layer on MI transition, which in turn can drastically change transport properties.

In a series of earlier works [8] transport and magnetotransport properties of (BEDO-TTF)$_2$ReO$_4\cdot$H$_2$O were intensively studied. However, only several of them contained analysis of magnetic properties at high temperatures. Static magnetization measurements have not yet been carried out. Unusual magnetic properties were observed below phase transition Me-Me' at T=213 K. On the basis of resistivity and EPR measurements they were interpreted in terms of SDW. However, the interpretation is not consistent well with measurements of Hall resistivity and low temperature RSA. Despite of diamagnetic state of anion layer (BEDO-TTF)$_2$ReO$_4\cdot$H$_2$O is very interesting model system to study influence of weak interactions on macroscopic transport properties in the vicinity of phase transition. By using EPR and SQUID methods we analyze characteristic features of fine ordering state in anion layer of polyhedral radical-cation salts (BEDO-TTF)$_2$ReO$_4\cdot$H$_2$O and (BEDT-TTF)$_x$Dy(NO$_3$)$_z$. Here the last salt has not been studied elsewhere.

## II. SAMPLES AND EXPERIMENTAL TECHNIQUE

### 1. Samples and structure

Two conducting charge-transfer salts based on (bis(ethylenedithio) tetrathiafulvalene)=BEDT-TTF molecule [8,9] and its derivative (bis(ethylenedioxy)tetrathiafulvalene)=BEDO-TTF molecule [10] have been studied. BEDO-TTF molecule is an analogue of BEDT-TTF, where four lighter oxygen atoms replace outer sulfur ones [1].

The crystals of (BEDO-TTF)$_2$ReO$_4\cdot$H$_2$O have been synthesized by electrochemical oxidation of BEDO-TTF in 1,2-dichloroethane-ethanol (10 V %) medium in constant current regime, I=0,3 µA, at the temperature 20° C. (Bu$_4$N)ReO$_4$ salt has been used as an electrolyte [11]. The black shiny crystals with the plate-like shape were grown on a platinum electrode within 1÷2 weeks. Typical sizes of the single crystals were (1x0.3x0.03) mm$^3$.

The radical cation salt (BEDO-TTF)$_2$ReO$_4\cdot$H$_2$O has layered structure with alternating radical cation layers of β''-type and insulating anion sheets along c-axis. Through 2 Dimensional (2D) network of short S…S (3.346÷3.661 Å) and S…O (3.152÷3.208 Å) contacts the radical cations BEDO-TTF$^{+1/2}$ form conducting layers [11]. The anion layer consists of the ReO$_4$ tetrahedral anions and water molecules. Due to hydrogen bonds H$_2$O molecules stabilize the packing of ReO$_4$ anions. At room temperature the hydrogen bonds are formed between H$_2$O molecule and only one ReO$_4$ anion. The length of this bond is 2.59 Å. At low temperatures T<200 K re-orientation of ReO$_4$ takes place. Here neighboring tetrahedrons are linked via solvent molecule [12]. Each H$_2$O molecule is now bound with two oxygen atoms of ReO$_4$ neighbors. It is likely the polymeric chains are formed. However, the details of this topology are not clear. According structure studies the transition at T=203 K is interpreted as type I Me-Me' phase transition [11-13].

Polycrystals of (BEDT-TTF)$_x$Dy(NO$_3$)$_z$ has been synthesized by electrocrystalization of BEDT-TTF in chlorobenzene with 10 V% ethanol under galvanostatic conditions (I=0,1μA) at temperature 20°C. (Bu$_4$N)$_3$Dy(NO$_3$)$_6$·H$_2$O salt has been taken as an electrolyte. Thin strip-shaped crystals of typical dimensions (1x0.2x0.01) mm$^3$ were grown on platinum anode during 2÷3 weeks. Due to insufficient quality of individual single crystals complete X-ray analysis has not been carried out yet. The microprobe data indicates that this salt contains Dy atoms in the structure.

## 2. Experimental methods and techniques

The standard CW X-band EPR method has been used to study magnetic and relaxation properties within 100<T<350 K (nitrogen gas flow cryostat). Set of small crystals was placed on a mica substrate in the TE$_{102}$ cavity. MW power level was kept as 2 mW. The EPR spectra were recorded during slow cooling down to 100 K and then during heating of the sample up to 350 K with typical rate 1÷2 K/min. Signal intensity, I$_{epr}$, has been determined by double integration (Schumakher-Slichter technique [14]) of EPR spectrum at magnetic field sweep δB$_{sw}$>5ΔB, where ΔB is peak-to-peak EPR linewidth. For single Lorentz EPR line this procedure gives ~10% of relative error. Absolute values of integrated intensity, I$_{epr}$, and g-factor were calibrated by using coal-pitch probe and CuSO$_4$·5H$_2$O.

The absolute values of spin contribution to magnetic susceptibility, χ$_s$, were measured independently by EPR as double integrated intensity, I$_{epr}$(T), and by SQUID as static susceptibility, χ$_p$(T). Therefore, further in the text the behavior of spin contribution to magnetic susceptibility, χ$_s$, extracted from EPR is discussed in terms of relative signal intensity I$_{epr}$ (χ$_s$~I$_{epr}$) scaled in arbitrary units.

Static paramagnetic susceptibility, χ$_p$, and magnetization, M, have been measured by using "Quantum Design" MPMS SQUID magnetometer at temperatures 2÷300 K and magnetic fields, B$_0$, up to 5 T. The randomly oriented powder of (BEDO-TTF)$_2$ReO$_4$·H$_2$O and several aligned single crystals of (BEDT-TTF)$_x$Dy(NO$_3$)$_z$ were put on a mica substrate oriented perpendicularly to the external magnetic field B$_0$. Typical specimen taken for a measurement was of 5÷15 mg. Diamagnetic contribution of mica substrate measured at 300 K was less than 1% of paramagnetic signal. Diamagnetic response measured within entire temperature range was subtracted from basic SQUID signal.

Temperature dependences of paramagnetic susceptibility were measured at B$_0$=1 T in order to obtain reliable signal from small masses of organic conductor. For (BEDT-TTF)$_x$Dy(NO$_3$)$_z$ the sample consisted of several individual micro crystals. Magnetization measurements were performed after cooling the samples in the field 1 T down to 2 K. Then at 2 K the field B$_0$ was increased to 5 T to make sure of absence of ferromagnetic impurities. Field dependences were measured by decreasing magnetic field from 5 T to 0.

## III. EXPERIMENTAL RESULTS

### 1. EPR in (BEDO-TTF)$_2$ReO$_4$·H$_2$O

Symmetric EPR signal has been observed within studied temperature range 100÷300 K. The EPR linewidth at 300 K reaches 45 G, which is a typical value for the BEDO-based organic metals [1]. The absence of HFS quartet line typical for 5d$^3$ spins of Re$^{4+}$ indicates that all metal cations are diamagnetic Re$^{7+}$ with 5d$^0$ configuration. The comparison of resistivity data, ρ(T), and EPR linewidth, ΔB(T), in (BEDO-TTF)$_2$ReO$_4$·H$_2$O are shown in Fig. 1. As ρ(T) reveals metallic behavior, ΔB(T) also gradually narrows at cooling. Metallic behavior of resisitivity is observed down to 4 K. In the vicinity of T=203 K both parameters undergo abrupt changes: resistivity drops from 0,21 Ω cm to 0,14 Ω cm, which is ~30%, and the linewidth increases from 32,37 G to 36,35 G, which is ~12%. Hence, the changes of magnetic and transport properties at 203 K can be interpreted as phase transition. Meanwhile, broadening of EPR signal below the transition disagrees with earlier published data [13]. The temperature evolution of both ρ(T) and ΔB(T) has irreversible (histeresis) behavior at cooling and heating. The width of thermal histeresis for ΔB(T) is ΔT=12 K, giving the estimate of transition temperature T=203 K. The linewidth, ΔB(T), is proportional to the temperature with the same coefficient d(ΔB(T))/dT=0,15 G/K for 100<T<203 K and 203<T<300 K segments (see fitting lines in Fig. 1). The resistivity only shows pronounced slope with dρ(T)/dT=1,4·10$^{-3}$ Ω cm/K below 215 K. The transition takes place at T= 223 K (half width). The temperature dependence of g-factor shown in Fig. 2 also indicates the transition at 203 K, although revealing very slight change of Δg=0,0004 near transition. At T<203 K g-factor is temperature independent and slightly increases at T>203 K (see fitting lines on Fig.

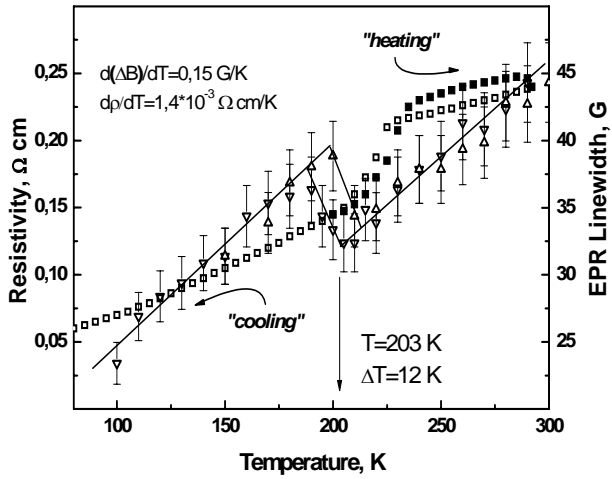

FIG.1 Temperature dependences of EPR linewidth at "heating" (Δ) and "cooling" (∇) for (BEDO-TTF)$_2$ReO$_4$·H$_2$O and of the dc resistivity, ρ(300 K)≈0,25 Ω cm.

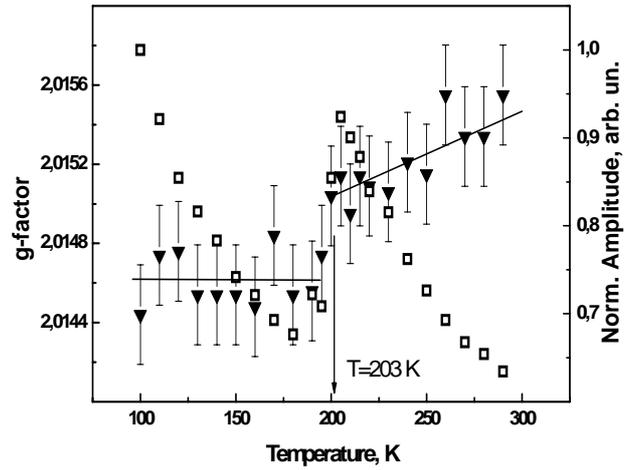

FIG. 2 Temperature dependences of g-factor at cooling and normalized amplitude of the EPR signal for (BEDO-TTF)$_2$ReO$_4$·H$_2$O.

2). The absolute values g(300 K)=2,015(6) and g(100 K)=2,014(4) are close to g-factor of free electron. The behavior of EPR linewidth and g-factor is consistent with metallic state above and below the transition temperature T=203 K. Therefore, it can be interpreted as Me-Me' phase transition, which is in agreement with earlier published results [11-13].

The spin contribution to the magnetic susceptibility has been obtained by double integration of EPR spectra. The temperature dependence of the integrated intensity, I$_{epr}$(T), is shown in Fig. 3 (see onset). The solid line indicates the best fit, where the $I_{epr}$ is linearly proportional to the temperature within the range 100÷300 K. In this range the intensity proportionally decreases nearly 2 times at cooling. Although, the amplitude of EPR signal clearly reveals the transition (see Fig. 2) the intensity and, respectively, spin susceptibility does not show any changes at 203 K. This behavior does not agree with [13], where EPR intensity showed 2 times drop at 213 K. One of the possible reasons is that EPR signal corresponds to incomplete response of spin ensemble. To confirm this result the measurements of static magnetic susceptibility has been carried out.

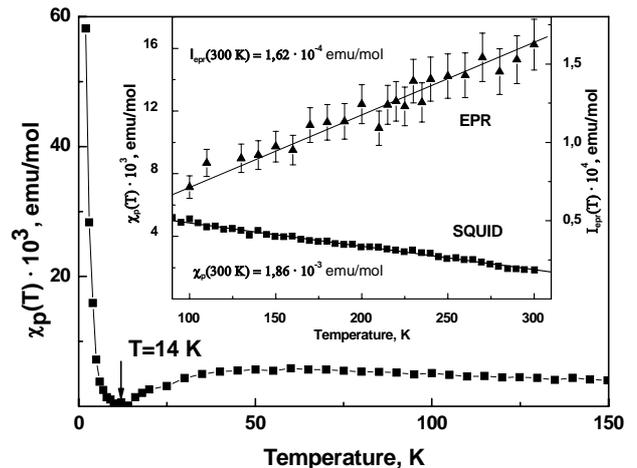

FIG.3 Temperature dependence of paramagnetic susceptibility χ$_p$ for (BEDO-TTF)$_2$ReO$_4$·H$_2$O. The inset shows high temperature segments of χ$_p$(T) and double integrated EPR intensity, I$_{epr}$(T). Solid lines are approximations.

## 2. SQUID measurements in (BEDO-TTF)$_2$ReO$_4$·H$_2$O

The solid squares in Fig. 3 denote the temperature dependence of the paramagnetic susceptibility, $\chi_p(T)$, measured by SQUID magnetometer. At 300 K the absolute value of $\chi_p$ is 1,86·10$^{-3}$ emu/mol. It increases up to 5,07·10$^{-3}$ emu/mol at 100 K (see onset on Fig. 3). Within the range 100 K<T<300 K $\chi_p(T)$ obeys quasi-linear dependence (solid line on the onset) with K=1,43·10$^{-5}$ emu·K/mol. At 14<T<40 K the $\chi_p$ decreases following the trend $\chi_p \to 0$ at T→14 K (see Fig. 3). This behavior contradicts with earlier reported SDW state bellow 35 K [13]. Below 14 K $\chi_p$ quickly recovers to 6·10$^{-2}$ emu/mol, which is consistent with Curie-Weiss estimate. In agreement with our EPR results the paramagnetic susceptibility is not sensitive to the transition at 203 K. However, low temperature behavior shown in Fig. 3 indicates that the total spin system consists of the localized moments with strong antiferromagnetic correlations. In contrast to EPR intensity, $I_{epr}$, the paramagnetic susceptibility, $\chi_p$, takes into account the total magnetic response of the whole spin ensemble. Thus, the $I_{epr}(T)$ reflects the evolution of a fraction of spins contributing to the EPR signal. We assume that this subsystem is related to delocalized charge carries (π-type holes). Hence, the temperature dependence can be attributed to the evolution of density of states, $n(\varepsilon_F)$. It is shown in [13], that there is no maximum at $\varepsilon_F$ and $n(\varepsilon_F)$ ~6 DOS/(unit cell), which is relatively weak compare to β-type metals. So, the fraction of delocalized spins can form a minor Pauli contribution to $\chi_p$. The linear temperature evolution of the $I_{epr}(T)$ indicates that the density of states at Fermi level decreases twice while cooling down to 100 K. The unit cell volume, V, determined from X-ray analysis [12] decreases from 2781,3 Å$^3$ at 293 K to 2710,(0) Å$^3$ at 150 K. Hence, one could expect increasing of the EPR intensity at low temperatures. To satisfy the above conditions the effective concentration of delocalized spins has to decrease. Meanwhile, those spins "disappeared" for EPR still contribute to paramagnetic susceptibility $\chi_p$ measured by SQUID. The response of this fraction of the local moments would be dominating. Indeed, the estimates of absolute value of spin susceptibility $\chi_s \sim I_{epr}(300 K)$ is considerable less than $\chi_p(300 K)$. These results are in agreement with Hall resistivity data in [13]. Indeed, delocalized carriers are holes and localized moments are electrons. We have also observed that right after SQUID measurements, which typically take hours in cryostat chamber, the EPR parameters of the sample differ from the ones measured prior to SQUID experiments. Evidently, vacuum pumping and open helium atmosphere affect the water content in the crystal volume and respectively affect the phase transition [15].

## 3. EPR in (BEDT-TTF)$_x$Dy(NO$_3$)$_z$

The radical cation salt (BEDT-TTF)$_x$Dy(NO$_3$)$_z$ is characterized by single EPR line with axial anisotropy of g-factor, $g_\parallel$=2,004(1) and $g_\perp$=2,008(3) and ΔB=13÷17 G at 300 K. Various ΔB within the range are observed after temperature cycling. Narrow EPR line is in a good agreement with semiconducting behavior of dc resistivity, ρ(T), shown on Fig. 4. Similar to previous case this EPR signal can be attributed to the semiconducting radical cation sub-lattice of BEDT-TTF [1]. No other signals, which could be attributed to magnetic Dy$^{3+}$ ions, are observed. Temperature dependence ΔB(T) demonstrates irreversible behavior. At "cooling" regime (measurements while

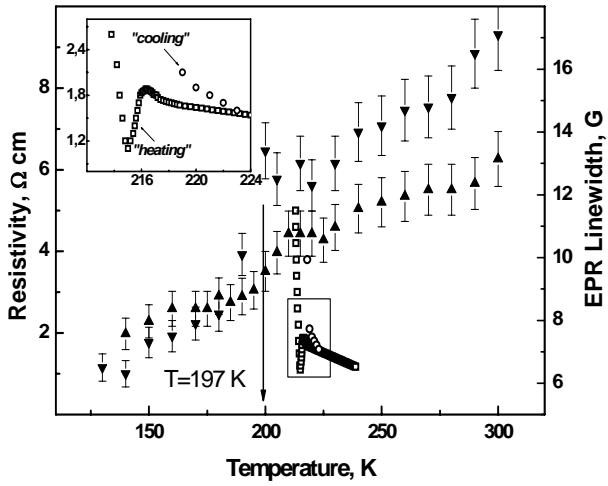# 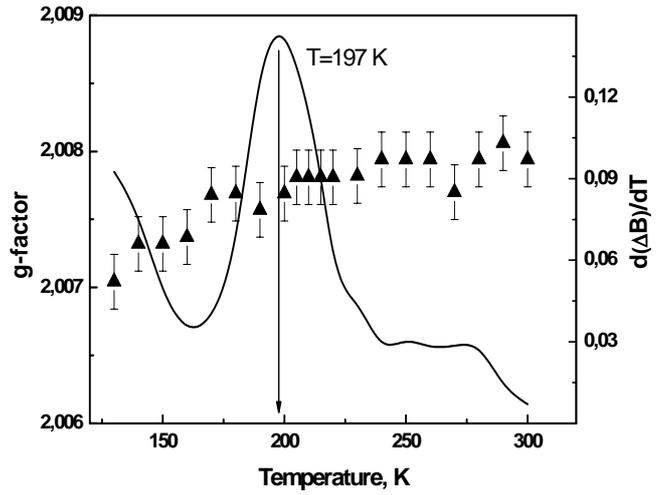

FIG.4 Temperature dependences of EPR linewidth at heating (▲) and cooling (▼) for $(BEDT-TTF)_2Dy(NO_3)_4$ and of dc resistivity. The inset shows detailed resistivity evolution in the vicinity of 200 K.

FIG. 5 Temperature dependences of g-factor at heating and temperature evolution of the first derivative of EPR linewidth, $d(\Delta B)/dT$, for $(BEDT-TTF)_2Dy(NO_3)_4$.

cooling, see down triangles on Fig. 4) $\Delta B$ gradually decreases from 17,2 G to 6,3 G at 130 K. In the vicinity of 200 K $\Delta B$ drops from 13,5 to 8,9 G within $\Delta T=10$ K. At "heating" (up triangles) from 130 K to 195 K $\Delta B(T)$ follows the same dependence. In the range $\Delta T=20$ K near 197 K $\Delta B(T)$ increases from 8,1 to 10,6 G, which is about 2 times less than 4,6 G fall at "cooling". In the range $215<T<300$ K the linewidth increases more slowly reaching 13,1 G at 300 K. Irreversible behavior of $\Delta B(T)$ correlates with temperature hysteresis of $\rho(T)$ near 216 K shown on the onset to Fig. 4. Thus, both resistivity and EPR linewidth distinguish states of BEDT-TTF sublattice before and after cooling below 197 K. With the time $\Delta B$ approaches initial values of ~18 G. In "heating" regime the transition at 197 K is clearly seen as depicted on Fig. 5 for the first derivative curve, $d(\Delta B)/dT$, where the initial curve $\Delta B(T)$ was taken from smooth interpolation of respective experimental data. The temperature evolution of g-factor obtained at "cooling" and "heating" does not show any sharp changes at 197 K (see Fig. 5, "heating" regime). Above 130 K the g-value gradually increases from $g=2,007(0)$ up to $g=2,008(0)$ at 300 K. Similar to $(BEDO-TTF)_2ReO_4$ salt the changes of transport and relaxation properties at 197 K can be attributed to structural transition.

## 4. Magnetic susceptibility in $(BEDT-TTF)_xDy(NO_3)_z$

Spin contribution to paramagnetic susceptibility of BEDT-TTF layers measured by double integration of EPR signal is shown on Fig. 6 ("cooling" regime, down triangles) as reciprocal intensity, $I_{epr}^{-1}(T)$. Below 197 K the intensity $I_{epr}$ sharply increases reaching $I_{epr}(170\ K)/I_{epr}(210\ K)=2,6$. Reciprocal paramagnetic susceptibility, $\chi_p^{-1}$, is also shown in Fig. 6 for perpendicular orientation of texturated sample $(BEDT-TTF)_xDy(NO_3)_z$ to the external field $\mathbf{B}_0$. At $T<250$ K $\chi_p^{-1}(T)$ obeys Curie–Weiss law (solid line) with $C=13,00$ emu·K/mol, $\theta=-0,08$ K. Strong angular dependences of $\chi_p$ has been observed. These results are subject of separate paper. The observed dependencies are interpreted in terms of the anisotropy of ligand field and strong spin-orbit coupling for $4f^9$ electrons of $Dy^{3+}$ [16]. However, due to incomplete alignment of microcrystals (sample $(BEDT-TTF)_xDy(NO_3)_z$ is not a single crystal) the full anisotropy was not resolved. Meanwhile, for both orientations in the range 100÷300 K no sharp changes, which can be attributed to a transition, are found. Effective magnetic moment, $\mu_{eff}$, extracted from $\chi_p^{-1}(T)$ consist $\mu_{eff}/\mu_B= 10,19$.

Field dependence of static magnetization M is described by Brillouin function with respective moment J=16/2. The magnetic moment of Dy$^{3+}$ is J= L+S=5+5/2=15/2 [16] and S=1/2 may come from BEDT-TTF$^+$. Hence, the paramagnetic response of (BEDT-TTF)$_x$Dy(NO$_3$)$_z$ measured by SQUID is consistent with calculated total moment 15/2 for Dy$^{3+}$. As soon as the Dy$^{3+}$ ions in Dy(NO$_3$)$_z$ layers give major contribution to paramagnetic response they determine respectively the temperature behavior $\chi_p(T)$. One of possible anion stoichiometry is Dy(NO$_3$)$_4$, where Dy$^{3+}$ is coordinated by 8 oxygen's. Hence, BEDT-TTF$_x$ has the charge +1.

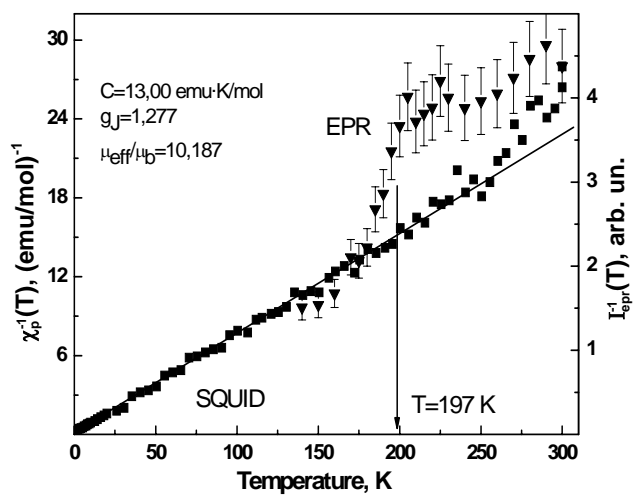

FIG.6 Temperature dependences of reciprocal paramagnetic susceptibility $\chi^{-1}_p$ ($\mathbf{B}_0 \perp c$) for (BEDT-TTF)$_2$Dy(NO$_3$)$_4$ and double integrated EPR intensity $I^{-1}_{epr}$. Solid line is Curie-Weiss approximation with parameters indicated on the plot.

As soon as $\chi_p(T)$ is dominated by non-interacting Dy$^{3+}$ moments the spin contribution BEDT-TTF is negligible. Indeed, the abrupt changes at 197 K visible on EPR are not seen in SQUID measurements. The magnetic system of Dy$^{3+}$ ions is not sensitive to Dy(NO$_3$)$_4$ configuration. In turn, the structural rearrangement in conducting BEDT-TTF layers does not cause a noticeable effect on magnetism of Dy$^{3+}$. Thus, the transition at 197 K is not induced or at least influenced by the magnetic state of anion complexes.

## IV. DISCUSSION

### 1. Magnetic transformations in (BEDO-TTF)$_2$ReO$_4 \cdot$H$_2$O

Metal-insulator transition at temperatures T≥80 K is a common feature for most of BEDT-TTF-based charge-transfer salts with tetrahedral and octahedral anions [1]. Independently on electronic band characteristics of 2D S..S network the transition shifts to higher temperatures as polymeric chain or solvent hydrogen bonds put a link between individual anions. For d-metal complexes with magnetic ions the transition temperature may raise to ~200 K or the compound becomes a semiconductor. The EPR signal in the BEDT-TTF salts with the metal-complex anions [M(Hal)$_4$] (where M is magnetic ions, M= Cu$^{2+}$, Mn$^{2+}$, Fe$^{3+}$ and Hal=Cl, Br) undergoes dramatic broadening from 50-120 G and 220 G for the first two salts respectively up to 700 and 1400 G for the compounds with Fe$^{3+}$ in anion sublattice [17]. For these compounds transport and magnetic properties at high temperatures are interpreted in terms of structural transformations in anion sublattice. Here, the magnetic state of metal ions undoubtedly plays the important role.

Despite significant difference in the structure of conducting layer, the salts (BEDO-TTF)$_2$ReO$_4 \cdot$H$_2$O and (BEDT-TTF)$_x$Dy(NO$_3$)$_z$ demonstrate similarities in the vicinity of 200 K. The phase transition in both compounds is induced by re-arrangement of anions without drastic changes of quasi-electronic spectra. The network of hydrogen bonds in anion layer and the absence of sulfur atoms in outer rings of BEDO-TTF molecule diminish the influence of anions re-orientation on band characteristics in comparison with BEDT-TTF. Partially this is because the sulfur atoms in inner rings are less sensitive to the order/disorder in the terminal ethylene groups of BEDO-TTF molecule. Note that (BEDT-TTF)$_2$ReO$_4$, which does not contain a H$_2$O "bridge" between ReO$_4$ demonstrates the metal-insulator transition at 80-90 K [18].

The increase of $\Delta B$ (~30 %, Fig. 1) at T≈203 K is in a good agreement with lower resistivity below the transition. The absence of transition on $\chi_p$(T) curve is consistent with continuous compression of unit cell volume at cooling [11]. However, our data disagree with earlier published results [13], where ΔB(T) drops ~25% at 213 K and stays temperature independent down to 90 K. Moreover, EPR intensity, determined as I~A(ΔB)$^2$ (A- signal amplitude) also drops 50% at T=213 K. Thus, for the same system (BEDO-TTF)$_2$ReO$_4$·H$_2$O at least two different types of physical properties can be observed. For the crystals showing metallic behavior of resistivity down to helium temperatures (type I, current study) Me-Me' transition does not reveal itself on $\chi_p$(T) and EPR linewidth increases below the transition. Maximum $\chi_p$(T) on Fig. 3 indicates antiferromagnetic interactions [19]. Meanwhile, at T<14 K paramagnetism quickly recovers. In contrast to [13] magnetic transformations at these temperatures do not influence on metallic behavior of resistivity. At T<2 K the samples are superconductors. The detailed study will be subject of separate paper. The crystals of type II [13] also show superconductivity but have more complicated phase diagram. They are characterized by temperature independent EPR linewidth at 90<T<213 K / 213<T<300 K with ~25% drop at 213 K and metallic behavior of ρ(T) down to 35 K. EPR intensity is temperature independent within 90<T<213 K and decreases below 90 K. Meantime, at T<35 K ΔB decreases while ρ(T) becomes semiconducting. The Hall resistivity reveals anomalies at ~10, 35 and 50 K. Low temperature EPR and transport properties of type II crystals were described in terms of SDW state.

We attribute the above variety of properties to the H$_2$O distribution in anion layers. Due to migration of H$_2$O molecules within anion layer 1D chains and/or 2D inhomogeneous "clusters" of ReO$_4$ can appear below transition at 200 K. Respectively, BEDO-TTF sublattice will tend to either form alternating metallic and dimerized insulating cation layers (type I) or stay uniform with further SDW state at low temperatures (type II). In case of H$_2$O deficiency the phase separation with co-existing stoichiometries (BEDO-TTF)$_2$ReO$_4$ (**BEDO-1**) and (BEDO-TTF)$_2$ReO$_4$·H$_2$O (**BEDO-2**) can take place.

From [18] one could expect semiconducting properties below phase transition in **BEDO-1**, where M-I transition expected to happen at ~90 K. Evidently, this phase can also be responsible for low temperature properties of type II crystals. The question arise whether amplitude, A(T), and ΔB are related to the same group of spins? At T>90 K both phases contribute to metallic behavior of resistivity, whereas EPR linewidth is mostly determined by **BEDO-1**. Indeed as the EPR linewidth for **BEDO-1** is temperature independent the behavior of intensity will be determined by the amplitude, A(T), of the total signal. Signal from **BEDO-2**, which is broader than that from **BEDO-1**, due to higher conductivity also contribute to the amplitude. Thus, depending on the fraction with stoichiometry **BEDO-2** the transition can appear on the temperature dependence of EPR amplitude (see Fig. 2). Hence one can observe the phase transition on I(T)~A(ΔB)$^2$ curve. However, double integrated EPR intensity or paramagnetic susceptibility as depicted on Fig. 3 do not show any abrupt changes. One more evidence in favor of two-phase interpretation is different temperature behaviors of $\chi^{-1}_p$(T) and $I^{-1}_{epr}$(T) shown in the same Figure 3. Fraction of spins contributing to EPR decreases at cooling while total paramagnetic susceptibility is well described as ensemble of localized spins with antiferromagnetic correlations over entire temperature range.

We assume that the crystals of (BEDO-TTF)$_2$ReO$_4$·H$_2$O studied in current work correspond to type I crystals, where re-orientation of ReO$_4$ leads to internal inhomogeneity. Below the transition the alternating of cation layers takes place. Certain layers stay metallic while other ones form antiferromagnetic dimmers. We assume that, in layered conductors based on BEDO-TTF with tetrahedral counter ions weak interactions (hydrogen bonds) in anion layer may lead to the changes of EPR and transport properties of the whole crystal at relatively high temperatures ~200 K.

## 2. Magnetic transformations in (BEDT-TTF)$_2$Dy(NO$_3$)$_4$

Due to similar layered structure the transition in (BEDT-TTF)$_x$Dy(NO$_3$)$_z$ detected by EPR at T=197 K may also indicate re-arrangement in Dy(NO$_3$)$_z$ layers. From above assumptions most probable stoichiometric formula is (BEDT-TTF)$_2$Dy(NO$_3$)$_4$. Field dependence of magnetization confirms presence of Dy$^{3+}$ in the compound and works in favor of the stoichiometry with x=2 and z=4. As a rule Dy$^{3+}$ cations are coordinated by two oxygen atoms of each NO$_3$ group [20]. Outer N-O group can also form hydrogen bonds with terminal ethylene groups of BEDT-TTF molecule. Therefore, re-arrangement of Dy(NO$_3$)$_4$ would affect S..S network of conducting layer. Irreversible resistivity shown on the inset in Fig.4 is consistent with this statement. On "heating" the dependence reveals partially a new state, which is different from the one observed at "cooling". The only property that could be changed between "cooling" and "heating" procedures is the amount of disorder in Dy(NO$_3$)$_4$ layers. Thus, the individual anions do not behave as independent complexes. One can suppose that several metastable configurations with different relative arrangement of NO$_3$ groups take place at high temperatures. Due to hydrogen bonds with NO the ethylene groups in BEDT-TTF sublattice experiences disorder. Below the transition T<197 K anion complexes occupy new configuration which is more uniform. Hence, BEDT-TTF sublattice is influenced by less amount of disorder. Indeed, Fig.4 shows that ΔB(T) has more gradual evolution compare to sharp changers of ρ(T). At "heating" above 197 K configurations with higher energies are not occupied. Therefore, the lower segment of ρ(T) hysteretic curve is related to thermodynamically stable state. Respectively EPR linewidth is smaller. The dependences $\chi^{-1}_p(T)$ and $I^{-1}_{epr}(T)$ in Fig.6 demonstrate mutual evolution of J=15/2 spin of Dy$^{3+}$ and S=1/2 of BEDT-TTF$^+$. Sharp increase of spin susceptibility of cation layers may be attributed to chemical compression of unit cell, which is most likely the characteristic feature of radical cation salts with polyhedral anions. The magnetic response of Dy$^{3+}$ moments depicted in Fig.6 does not sense anion re-arrangement. Indeed, due to shielding effect the influence of ligand field on 4f-electron spins is very weak, so that spin-orbit coupling is not changed during re-arrangement. Since localized f-orbitals do not contribute to hybridization the influence of Dy$^{3+}$ local moments on spin system of conducting layer is negligible. Hence, relaxation channel to magnetic ions reservoir can only be realized through the dipole-dipole interactions. The distances between Dy$^{3+}$ and nearest short S..S contacts in conducting layer are too long (d>10 Å) to observe this effect in the studied temperature range. Thus, narrow EPR signal in (BEDT-TTF)$_x$Dy(NO$_3$)$_4$ is well understood because the magnetic f-elements do not induce paramagnetic relaxation in contrast to d-metals. Evidently, magnetic superexchange between neighboring Dy ions is also too weak to form coupled states at T>2 K.

## V. CONCLUSION

In this work we have shown that EPR and SQUID provide consistent information about spin ensemble of quasi-2D organic conductors. It is concluded that sub-system of delocalized spins in BEDO-TTF layer contribute to EPR signal of (BEDO-TTF)$_2$ReO$_4$·H$_2$O, whereas static magnetic susceptibility measured by SQUID is mainly governed by localized spin states. The relative proportion between two sub-systems is determined by distribution of solvent molecules H$_2$O in anion layer. It is concluded while analyzing EPR intensity, $I_{epr}(T)$, that the concentration of delocalized spins is not constant and gradually decreases with cooling. Thus, the electronic phase transition Me-Me' result in alternating metallic and semiconducting layers in BEDO-TTF sublattice. Due to inhomogeneous solvent content or its peculiar topology in the layer the crystals (BEDO-TTF)$_2$ReO$_4$·H$_2$O reveal various transport and magnetic properties at low temperatures. At phase transition the delocalized spins contributing to EPR signal does not reveal strong changes. However, their relaxation rate is sensitive to the anion rearrangements. This behavior can be explained in terms of decreasing of amount of disorder in ethylene groups of BEDO-TTF molecules. In $(BEDT-TTF)_xDy(NO_3)_y$ the local spin moments with S=1/2 in the conducting BEDT-TTF layer contribute to EPR signal. However, the magnetic moments J=15/2 of $Dy^{3+}$ dominate in $\chi_p(T)$ measured by SQUID. Evidently, conductivity is induced by thermo activated hopping mechanism. Since the EPR intensity at T~197 K increases 2 times higher than $I_{epr}$(300 K) the low temperature phase can be attributed to paramagnetic insulator state.

The analysis of experimental data indicates that despite the peculiarities of individual TTF-based conductors with polyhedral anions their physical properties at high temperatures T>100 K are ruled by structural motif of anion layer. The arrangement of anions is very sensitive to weak interactions such as hydrogen bonds with solvent molecules and terminal ethylene groups of cation radical (BEDT-TTF or BEDO-TTF) and/or magnetic interactions. Re-orientations of polyhedrons lead to Me-Me' and Me-I phase transitions observed at T=203 K in $(BEDO-TTF)_2ReO_4·H_2O$ and at T=197 K in $(BEDT-TTF)_2Dy(NO_3)_4$ respectively. The insert of f-metal into individual anion complex does not lead to the magnetic effect on phase transition. It is explained by negligible hybridization with p-electrons of ligand. Thus, nor interlayer π-f interactions with delocalized π-type charge carriers (holes) nor intralayer f-f interactions among local magnetic moments of $Dy^{3+}$ exist at T>2 K in $(BEDT-TTF)_2Dy(NO_3)_4$.

We suggest that synthesis of bi-metallic anion structures based on d-metal complexes is prospective approach for the synthesis of new polyfunctional materials for molecular electronics. We also suggest that due to easy polarization of π-electrons in cation layer the metal chelates as counter ions are promising for creating of effective interlayer magnetic interactions.

## ACKNOWLEDGEMENTS

The authors are thankful to Dr. L.I. Buravov for resistivity data. This work was supported by the RFBR (grant No. 1380.2003.2), RFBR-DFG (grant No. 03-02-04023) and NWO 047.008.020.